\documentclass[twocolumn,nofootinbib,showpacs,prd]{revtex4}

\usepackage{amsmath}
\usepackage{amsfonts}
\usepackage{amssymb}
\usepackage{graphicx}

\begin{document}

\title{Holographic dark matter and dark energy with second order invariants}

\author{Alejandro Aviles${}^\S{}^\star$, Luca Bonanno${}^\P{}^\flat$, Orlando Luongo${}^\ddag{}^\dag{}^\S$ and Hernando Quevedo${}^\ddag{}^\S$}

\address{${}^\ddag$Dip. di Fisica and Icra, Universit\`a di Roma "La Sapienza", I-00185 Roma, Italy,}
\address{${}^\flat$Institut f$\ddot{u}$r Theoretische Physik, J. W. Goethe - Univ. Max-von-Laue-Stra${\ss}$e, 1 60438
Frankfurt am Main, Germany,}
\address{${}^\P$Dip. di Fisica, Universit\`a di Ferrara and INFN, via Saragat 1, 44100 Ferrara, Italy,}
\address{${}^\dag$Dip. di Scienze Fisiche, Universit\`a di Napoli ''Federico II'', Via Cinthia, Napoli, Italy,}
\address{${}^\S$Instituto de Ciencias Nucleares, Universidad Nacional Aut\'onoma de M\'exico, Mexico,}
\address{${}^\star$Depto. de Fisica, Instituto Nacional de Investigaciones Nucleares, Mexico.}

\begin{abstract}
One of the main goals of modern cosmology remains to summon up a
self consistent policy, able to explain, in the framework of the
Einstein's theory, the cosmic speed up and the presence of Dark
Matter in the Universe. Accordingly to the Holographic principle,
which postulates the existence of a minimal size of a physical
region, we argue, in this paper, that if this size exists for the
Universe and it is accrued from the independent geometrical second
order invariants, it would be possible to ensure a surprising source
for Dark Matter and a viable candidate for explaining the late
acceleration of the Universe. Along the work, we develop low
redshift tests, such as Supernovae Ia and kinematical analysis
complied by the use of Cosmography and we compare the outcomes with
higher redshift tests, such as CMB peak and anisotropy of the cosmic
power spectrum. All the results indicate that the models presented
here can be interpreted as unified models that are capable to
describe both the dark matter and the dark energy.
\end{abstract}


\maketitle

\section{Introduction}

General Relativity has been deemed as one of the cornerstones of
modern theoretical physics. All its predictions, especially in the
Solar System regime, found numerous experimental evidences
\cite{cliff}. Despite the theory promises to completely clear up the
dynamical properties of the whole Universe, many attempts spent to
explain those properties, at cosmological scales, failed to be
predictive. The main flaw arose when in 1998 it was first discovered
\cite{SNeIa1,SNeIa2}, that a late positive acceleration dominates
the dynamics of the Universe.

This key feature was straight observed by the use of Supernovae Ia,
as distance indicators and later confirmed by various experimental
evidences \cite{sn3}; however, it was also clear that if one takes
into account only the baryonic and (cold) dark matter (DM), as
gravitational sources, GR does not predict an accelerated scenario,
as expected according to observations.

Thus, it has been argued that a further new ingredient should be
enclosed within the cosmological puzzle \cite{coppa}. Its physical
nature remains undiscovered and so, due to the lack of a
self-consistent explanation of it, we refer to this missing
counterpart as Dark Energy (DE), which might counterbalance the
gravitational effects. Moreover, quite surprisingly, observations
spilled out definitively that about 70$\%$ of the Universe is filled
by this unknown ingredient and in addition that about the 25$\%$ is
composed by DM. Therefore, numerous approaches have followed one
another, during the time, in order to elucidate the nature of these
unknown components.

At a first glance, being as the likely simplest explanation of the
acceleration, a cosmological constant term $\Lambda$ should
characterize the form of DE \cite{lam}. Unfortunately, its
consequent model, namely $\Lambda$CDM, undergoes several theoretical
issues, leading to the well known fine-tuning and coincidence
problems \cite{tsu}. For overcoming these issues many alternatives
to DE have been proposed \cite{nu}.

Without going into details on the various frameworks propounded in
literature, we focus on one intriguing task, which is represented by
the so-called Holographic principle (HP) \cite{hol}. The basic idea
lies on postulating that the maximum entropy inside a physical
region is not extensive, since it grows as the area of the surface.

By extending this postulate to cosmology, it would be feasible to
infer that the density of DE, namely $\rho_X$, should be
proportional to an infrared (IR) cut-off scale, namely $L$, and it
can be rewritten as

\begin{equation}\label{js}
\rho_X\propto \frac{1}{L^2}\,.
\end{equation}

The idea behind the cut-off scale is that a minimal information
should exist and the consistent density associated to this minimal
counterpart should be employed as an energy density. The latter one
may be therefore included within the Einstein's equations, in order
to provide the positive acceleration. Thereby, the problem of
understanding the nature of DE is shifted to the crucial issue of
determining $L$. An amusing loophole leads to the choice that a
viable $L$ is that one, able to account both DM and DE effects.

Many different cut-off scales have been examined during the last
years \cite{bingo,uhmamma2,uhmamma3}. In particular, in a recent
work it has been postulated that $L$ may be proportional to the root
mean square of second order geometrical invariants \cite{mio}. The
choice of a geometrical IR cutoff has been demanded in order to work
out the problem of causality, portrayed in \cite{cai}, by allowing
to solve it naturally, with only considering the form of space-time
in the framework of GR. The physical purport of a geometrical choice
of $L$ deals with the possibility that the geometry can be
considered as a self-accelerated source, endowing a DE term.

The main intent of this paper is to testify that the HP, with the
scale length proportional to curvature invariants, works fairly,
providing encouraging results for attesting that it should be
possible to regard to geometry as sources of DE and predictable DM.
As a first glance, then, our model can be considered as an unified
paradigm for describing either the dynamical properties at
cosmological scales or the presence of DM in the Universe.

The paper is then organized as follows; in Sec. 2 we describe the
theoretical features of our picture, in Sec. 3 we study and we
extend the work of \cite{mio} by performing cosmological tests on
the model. Sec. 4 deals with the use of the so-called Cosmography,
as a tool for discriminating the kinematics of the Universe, in the
case of our approach. Finally Sec. 5 develops the comparison of our
proposal with the anisotropies of the cosmic power spectrum. Sec. 6
is devoted to conclusions and perspectives.

\section{The theoretical framework}
\label{sec2}

In this section, we investigate the consequences following from the
assumption that an infrared cutoff exists that is proportional to
the independent second order invariants.

As in Ref. \cite{mio}, we invoke the HP as a way to solve the DE
paradigm, by relating the second order curvature (independent)
invariants to the DE density, in the way

\begin{equation}\label{cnn}
\rho_X=\frac{3\alpha}{8\pi G}\sqrt{|I_i|}\,;
\end{equation}
where with $I_i$ we express the generic invariant, while $\alpha$ is
a dimensionless constant. The above equation is written in this
form, since second order invariants are proportional to the inverse
fourth power of the IR cutoff, then for dimensional requirements, we
need the root mean square in the above form.

As pointed out in Ref. \cite{capozzox}, among the 14 curvature
scalar invariants, the most interesting ones are the Kretschmann,
Chern-Pontryagin and Euler invariants. Their forms are summarized as
follows
\begin{eqnarray}
I_1 &=& R_{\alpha\beta\gamma\delta}R^{\alpha\beta\gamma\delta}\,, \nonumber \\
I_2 &=& [{}^*R{}]_{\alpha\beta\gamma\delta}R^{\alpha\beta\gamma\delta}\,, \nonumber \\
I_3 &=&
[{}^*R{}^*]{}_{\alpha\beta\gamma\delta}R^{\alpha\beta\gamma\delta}\,,
\end{eqnarray}
where the stars indicate the corresponding dual counterparts. From
the first Matt\'e-decomposition of the Weyl tensor, it is easy to
get \cite{roberts}
\begin{eqnarray}
R_{\alpha\beta\gamma\delta}&=&C_{\alpha\beta\gamma\delta}+\frac{1}{2}
\big( g_{\alpha \gamma} R_{\beta \delta}-\nonumber\\
&&-g_{\beta \gamma} R_{\alpha \delta}- g_{\alpha \delta}R_{\beta \gamma}+ g_{\beta \delta}R_{\alpha \gamma}\big)\nonumber \\
&&-\frac{1}{6}( g_{\alpha \gamma}g_{\beta \delta}-g_{\alpha
\delta}g_{\beta \gamma})R\,.
\end{eqnarray}

Therefore, $I_1$, $I_2$ and $I_3$ can be expressed as follows
\cite{gendeb,Witten,Petrov,rinpen,carmi}
\begin{eqnarray}\label{KAPPA1}
I_1&=&C_{\alpha\beta\gamma\delta}C^{\alpha\beta\gamma\delta}+2R_{\alpha\beta}R^{\alpha\beta}-\frac{1}{3}R^2\,, \nonumber  \\
I_2&=&[{}^*C{}]_{\alpha\beta\gamma\delta}C^{\alpha\beta\gamma\delta}\,,\nonumber\\
I_3&=&-C_{\alpha\beta\gamma\delta}C^{\alpha\beta\gamma\delta}+2R_{\alpha\beta}R^{\alpha\beta}-\frac{2}{3}R^2= \nonumber \\
&=&-I_1+2R_{\alpha\beta}R^{\alpha\beta}-\frac{2}{3}R^2\,.\nonumber
\end{eqnarray}

From the above relations among invariants and tensors, one can get
the explicit expressions of the second order invariants, once the
space-time metric is known. We use hereafter a flat
Friedmann-Robertson-Walker cosmology
\begin{equation}\label{frw}
ds^2=dt^2-a(t)^2\big[dr^2+r^2(\sin^{2}\theta
d\theta^2+d\phi^2)\big]\,,
\end{equation}
and for future convenience it is also necessary to write down the
first Friedmann equation, being
\begin{equation}\label{F1bis}
\frac{H^2}{H_{0}^{2}}\equiv\left(\frac{\dot a}{a}\right)^2
\frac{1}{H_0^2}=\frac{8 \pi G}{3 H_0^2}\Big[\rho_X + \Omega_m
(1+z)^3 + \Omega_r (1+z)^4\Big]\,.
\end{equation}

Actually, the Hubble factor $H$ has been parameterized by $h$ as
$H_0 = 100 h \,\text{km}/\text{s}/\text{Mpc}$. The dimensionless
density parameters of matter and the relativistic components are
defined as $\Omega_m = 8 \pi G \rho^{(0)}_m/3 H_0^2\,$ and
$\,\Omega_r = 8 \pi G \rho^{(0)}_r/3 H_0^2$ respectively, where a
index $(0)$ denotes that the quantity under examination is evaluated
at $z=0$, with the scale factor $a(t)$ normalized to the unity
today, {\it $a(z=0)=1$}.

We obtain for the invariants the following expressions

\begin{eqnarray}\label{primo}
I_1&=&60\left\{(\dot
H+2H^2)^2+H^4\right\}\,,\nonumber\\
\,\nonumber\\
I_2&=&0\,,\nonumber\\
\,\nonumber\\
I_3&=& -12\Bigl\{5(\dot H+2H^2)^2+5H^4+\,\\
&+&2(\dot H+2H^2)H^2 \Bigr\}\,.\nonumber
\end{eqnarray}

By inserting the expressions for $I_1$ and $I_3$ (the only two
nontrivial invariants) into Eq. ($\ref{cnn}$) and by using the
Friedmann equation, we gain two differential equations, each one
providing the temporal evolution of the Hubble parameter. As in Ref.
\cite{mio}, we will refer to the cosmological models arising from
the invariants $I_1$ and $I_3$ as $mod_1$ and $mod_3$, respectively.
So, by keeping all these key features it will be possible in the
next sections to perform various experimental bounds on our two
models. Particulary, what is following deals with the limits over
their expansion history.

\section{The Expansion History}

In this section, we strengthen the robustness of the theoretical
assumptions developed in Sec. II, by investigating the expansion
history of the Universe for our models.

First of all, let us focus on the vacuum solutions of these models;
they refer to the settlement with $\Omega_r =0$ and $\Omega_m=0$.
This limiting case is important in order to investigate the
asymptotic behavior of our models\footnote{Moreover, one expects
that the vacuum solutions correspond to the manifolds with the
maximum number of symmetries; these manifolds can then be
interpreted as the background spaces on which the matter fields
evolve.}.

For this limiting case we can analytically solve Eq. ($\ref{F1bis}$)
and it turns out that the holographic fluid behaves as a perfect
fluid with constant barotropic factors
\begin{eqnarray}\label{euno}
w_{\text{eff}} &=& \frac{1}{3} -\frac{2}{3}\sqrt{-1 + 1/ (60 \alpha_1^2)}\,,\nonumber\\
\,\\
w_{\text{eff}} &=& \frac{7}{15} -\frac{2}{3}\sqrt{-24/25 + 1/ (60
\alpha_3^2)}\,,\nonumber
\end{eqnarray}
for mod$_1$ and mod$_3$, respectively. We can then choose the
constant $\alpha$ as
\begin{eqnarray}\label{nnso}
\alpha_1 = \frac{1}{\sqrt{60\times 5}}\,,\quad\quad \alpha_3 =
\frac{1}{\sqrt{29 \times 12}}\,,\nonumber
\end{eqnarray}
so that $w_{\text{eff}} =-1$ for both models. Clearly, these choices
lead to a de Sitter space-time as vacuum solutions and, as we will
show below, they will be in excellent agreement with observations.

\begin{figure}
\centering
\includegraphics[width=3.4in]{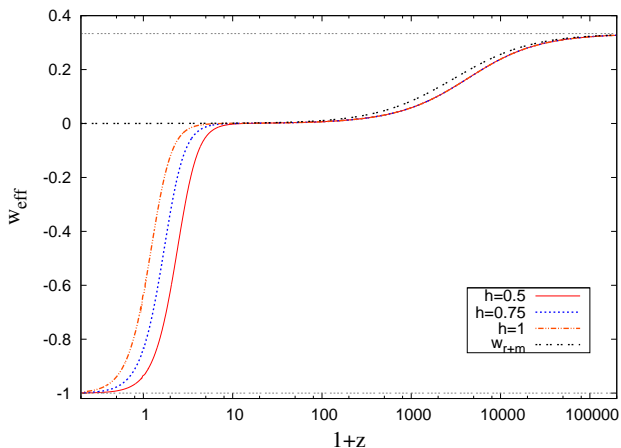}
\caption{(Color online). $w_{\text{eff}}$ for mod$_3$ with $\Omega_m
= 0.15$, $\Omega_r = 4.9 \times 10^{-5}$. The thick (red) line is
for  $h=1$; the dotted (blue) line is for $h=0.75$; dashed-dotted
(orange) line is for $h=0.5$. The double-dashed (black) line is the
effective fluid of joint matter and radiation, $w_{r+m}$. The
horizontal lines show the asymptotic behaviors
$w_{eff}\rightarrow1/3$ and $w_{eff}\rightarrow-1$} \label{fig1}
\end{figure}

In order to analyze the behavior of our models in the presence of
standard matter and radiation fields, we plot in Fig. \ref{fig1} the
corresponding EoS for different values of $h$ and arbitrarily chosen
values for $\Omega_m = 0.15$ and $\Omega_r = 4.9\times10^{-5}$.

However, to obtain the results reported in Fig. 1, it was necessary
to choose for $\Omega_m$ a value which is clearly lower than the
value obtained by WMAP 7. This is due to the fact that the HP also
appears as a source of the DM.

We can perceive that, until the epoch of matter-radiation equality
(approximately at $z \sim 3000$), the holographic fluid behaves as a
relativistic component with $w_{\text{eff}}=1/3$, afterwards it
behaves as a dustlike component, with $w_{\text{eff}}=0$, until it
passes a redshift threshold ($z \sim 10$) and begins to follow the
asymptotic value $w_{\text{eff}}\rightarrow-1$. Moreover, in Fig.
\ref{fig1} we plotted the EoS parameter of the joint matter and
radiation fluids, defined as

\begin{equation}
w_{r+m} = \frac{ \sum_{i} \rho_i w_i}{ \sum_{i} \rho_i} =
\frac{\Omega_r /3}{\Omega_r + \Omega_m (1+z)^{-1}}\,. \label{wrm}
\end{equation}

We can see that at high redshifts the behavior of the holographic
fluid resembles that of the source fields. However, it is
interesting to note that the effective EoS parameter departs from
$w=1/3$ before that of the matter-radiation fluid. As we shall see
in Sec. 5, this happens because the holographic fluids we consider
are more likely to mimic matter fluids than relativistic ones.

As first pointed out in the introduction, we suggest that both the
DE and DM counterparts may be explained by the choice of the IR
cutoff scale; so, to obtain a precise value for $\Omega_m$, we must
perform a chi-squared fit with the supernovae union 2 data set
\cite{Union2} and with the CMB shift parameter \cite{EstaBond}. The
combined test is sufficient to predict a specific value which will
be tested further in Sec. V, where we will compare the predictions
of our models with those of $\Lambda$CDM, by using CMB anisotropies.
All the remaining parameters, such as the Hubble constant,  will be
fixed by using the WMAP 7 maximum likelihood \cite{WMAP7y}.

The application of the standard definition of the CMB shift
\cite{EstaBond}
\begin{equation}\label{cmbsod}
\mathcal{R}=H_0 \sqrt{\Omega_m} \int^{z_{rec}}_{0}
\frac{dz}{H(z)}\,,
\end{equation}
presents some difficulties which can be overcome by using the
alternative definition \cite{Melch}
\begin{equation}\label{cmbshift}
\mathcal{R} \equiv 2 \frac{l_1}{l_1^{'}}\,,
\end{equation}
where $l_1$ is the position of the first peak on the CMB TT power
spectrum of the model under consideration, and $l_1^{'}$ is the
first peak in a flat FRW universe with $\Omega_m=1$. In particular,
for approaches providing a unified description of both DE and DM,
the latter expression must be used.

For any arbitrary model, $l_1$ is defined as
\begin{equation}
l_1 = \frac{D_A(z_{rec})}{s(z_{rec})}\,,
\end{equation}
where $D_A(z_{rec})$ is the comoving angular distance at
recombination, i.e.,
\begin{equation}
D_A(z_{rec}) = \int^{z_{rec}}_{0} (1+z) dz\,,
\end{equation}
and $s(z_{rec})$ denotes the sound horizon at recombination
\begin{equation}\label{shor}
s(z_{rec}) =  \int_{z_{rec}}^{\infty}\frac{c_s(z)}{H(z)}dz\,.
\end{equation}
Here  $c_s(z)$ is the sound speed of the photon-to-baryon fluid,
$c_s (z)= 3^{-1/2}(1+4\rho_b/3\rho_{\gamma})^{-1/2}$.

In the special case of $\Lambda$CDM, the shift parameter is
simplified to (\ref{cmbsod}) due to the fact that the cosmological
constant's contribution to the Hubble flow is negligible for
$z>z_{rec}$, so that it can be neglected in Eq. (\ref{shor}). In
general this approximation is not valid. This is the case of the
models that are the subject of this work. For these models, indeed,
at early times the holographic fluid mimics the whole background's
fluids. Therefore we will deal hereafter with the definition
(\ref{cmbshift}).

The best value for the CMB shift parameter inferred from the WMAP7yr
analysis is given by
\begin{equation}\label{yt}
\mathcal{R}=1.726 \pm 0.018\,.
\end{equation}
On the other hand, for the supernovae (SNe) fit we compare the
distance modulus
\begin{equation}\label{distmod}
\mu(z) = 25 + 5\log_{10} \left(\frac{d_L}{\text{Mpc}} \right)\,,
\end{equation}
with the observational data of the Union 2 data set which
encompasses 557 supernovae up to the redshift $z\sim$1.7, which
represents the highest value. Moreover, $d_L$ is the luminosity
distance defined as
\begin{equation}\label{lumdist}
d_L = c(1+z) \int^{z}_{0} \frac{dz}{H(z)}\,,
\end{equation}
and then via WMAP data we set the Hubble parameter today as
$h=0.704$.

\begin{figure}
\centering
\includegraphics[width=3.4in]{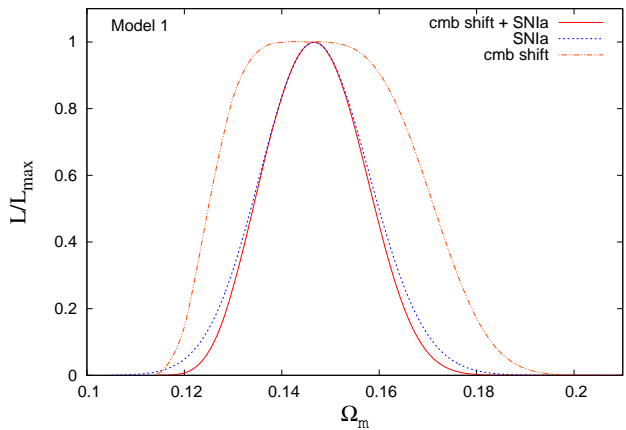}
\caption{(Color online). Mod$_1$ likelihood functions for the
expansion history analysis. The dot-dashed  (orange) line is for the
CMB shift analysis, the dashed (blue) line, for Union 2, and the
thick (red) line, for the joint Union 2 and CMB shift.} \label{fig2}
\end{figure}

\begin{figure}
\centering
\includegraphics[width=3.4in]{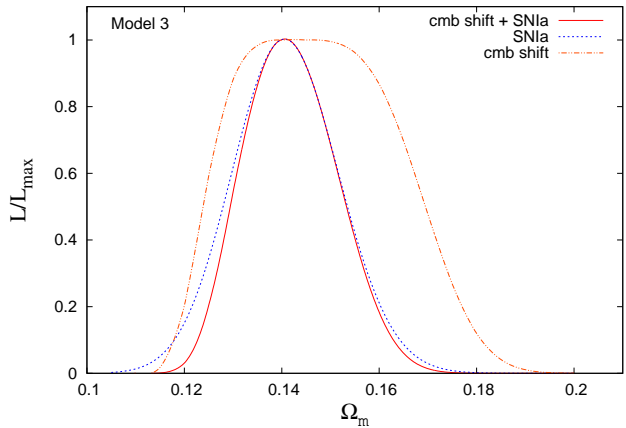}
\caption{(Color online). The same as in Fig. 2, but for mod$_3$.}
\label{fig3}
\end{figure}

The results are summarized in Table \ref{table1}. Figures \ref{fig2}
and Fig.\ref{fig3} show the likelihood functions of this analysis.

\begin{table}
\vspace{0.2cm}
\begin{tabular}{l|c|c|c}
\toprule %
 & Union 2  & CMB Shift &$\,\,$ Shift$+$Union2$\,\,$\\
&$\Omega_m\,\,\,$($\chi^2$) & $\Omega_m$ & $\Omega_m$ \\ \hline
Model 1 $\,\,$ & $\,\,\,$ $0.147^{+0.007}_{-0.007}$ $\,\,\,$ & $\,\,\,$ $0.145^{+0.020}_{-0.017}$ $\,\,\,$  & $\,\,$ $0.147^{+0.004}_{-0.004}$ \\
&$\,\,\,$ (543.5) $\,\,\,$ &$\,\,\,$ $\,\,\,$  & $\,\,$  \\ \hline
Model 3  $\,\,$ & $\,\,\,$  $0.141^{+0.007}_{-0.007}$ $\,\,\,$ & $\,\,\,$  $0.143^{+0.019}_{-0.017}$ $\,\,\,$  & $\,\,$ $0.142^{+0.006}_{-0.004}$\\
&$\,\,\,$ (543.6) $\,\,\,$& $\,\,\,$ $\,\,\,$  & $\,\,$  \\ \hline
\end{tabular}
\caption{Summary of the numerical results for $\Omega_m$. The quoted
errors shows the $68 \%$ confidence levels.} \label{table1}
\end{table}

\begin{figure}
\centering
\includegraphics[width=3.4in]{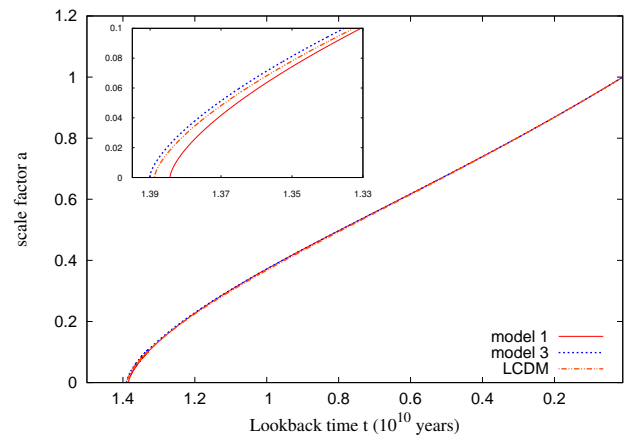}
\caption{(Color online). Expansion history of model 1 (thick line),
model 3 (dotted line), and the $\Lambda$CDM model (dot-dashed
line).} \label{fig4}
\end{figure}

Fig. \ref{fig4} shows, instead, the expansion history of the
Universe for the best fits previously discussed, and the scale
factor for the $\Lambda$CDM model with the best fit parameter values
given by WMAP 7 results \cite{WMAP7y}. We note that there are slight
differences that are more evident at early times. In other words,
according to the above results, our models are able to describe the
dynamics of the Universe by using a geometrical source for DM and
DE, with small differences with respect to the $\Lambda$CDM model at
late times.

\section{Low redshift regime: the Cosmography way}

Cosmography is an additional tool to test a given model.
Particularly, Cosmography represents a branch of cosmology, which
promises insight into the cosmological picture, for exploring the
kinematics of the Universe, without regards towards any a priori
specific model, postulated in the Friedmann equations.

The kinematics is very useful to understand the expansion history of
the Universe. First, as quoted, the kinematics does not depend on
the choice of the model, second the kinematics can be directly
fitted with the observable Universe, third it lies on the assumption
that the cosmological quantities should be expanded in series around
our time, $z=0$, giving rise to easygoing relations. In this sense,
naively, cosmography ingenuously builds on the simplest expediency
to investigate the universe's dynamics.

To apply cosmography one needs only the metric (5) that defines the
geometry of the universe. In other words, it does not take into
account any special modification of the Friedmann equations,
independent of the choice of the model.

This policy was first mooted from Weinberg's pioneering ideas. He
suggested expanding the scale factor in series of powers, giving the
possibility to relate it\footnote{Or the Hubble parameter, or the
luminosity distances and so on.} in terms of a Taylor series

\begin{eqnarray}\label{serie1}
a(t) & = &   1 + H_0 \Delta t - \frac{q_0}{2} H_0^2 \Delta t^2 +
\frac{j_0}{6} H_0^3 \Delta t^3 + \frac{s_0}{24} H_0^4
\Delta t^4 +\ldots\,,\nonumber\\
\end{eqnarray}
where it is worthwhile to get the positions

\begin{eqnarray}\label{pinza}
q &=& -\frac{1}{H^2} \frac{\ddot{a}}{a}\,, \nonumber\\
j &=& \frac{1}{H^3} \frac{a^{(3)}}{a}\,, \\
s &=& \frac{1}{H^4} \frac{a^{(4)}}{a}\,, \nonumber
\end{eqnarray}
which are known in the literature as the deceleration parameter, the
jerk parameter and the snap parameter, respectively. It is standard
convention to assume that Eqs. ($\ref{pinza}$) represent the
cosmographic set (CS), once each value has been evaluated at $z=0$.
Therefore, we commonly refer to the CS as the numerical values
assumed by the above coefficients of the Taylor expansion at late
time. Physically, this had to be argued since the series of $a(t)$
has been evaluated around our time. Therefore, the luminosity
distance $d_l$ can be rewritten as

\begin{eqnarray}\label{zump}
d_L & = & \frac{1}{H_0} \Bigl[ z + z^2 \Bigl(\frac{1}{2}-\frac{q_0}{2} \Bigr)+z^3\Bigl(-\frac{1}{6}-\frac{j_0}{6}+\frac{q_0}{6}+\frac{q_0^2}{2} \Bigr) + \nonumber\\
\,\\
& + & z^4 \Bigl(
\frac{1}{12}+\frac{5j_0}{24}-\frac{q_0}{12}+\frac{5 j_0 q_0}{12} -
\frac{5 q_0^2}{8}-\frac{5 q_0^3}{8} + \frac{s_0}{24}\Bigr)\Bigr]\,.
\nonumber
\end{eqnarray}

Afterwards these expansions, we have all the instruments needed to
perform a direct analysis. In fact, let us fit the Union 2 data
compilation by Eq. ($\ref{zump}$) in order to turn out experimental
limits over $q_0,j_0$ and $s_0$. Once the CS is known, it appears
easy to invert them and to relate the free parameters of our models
in terms of the CS. This guarantees experimental bounds on the
previous models, giving us the possibility to establish whether our
approaches work well or not in the low redshift regime.

\subsection{Experimental results}

In order to perform a more stringent check on the reliability of the
models, we test their predictions at low values of the redshift $z$,
by using the above results and those of Table I. In particular, we
interrelate the theoretical features of our models with the CS by
rewriting down $q,j,s$ in terms of the Hubble rate, i.e.

\begin{eqnarray}\label{parametri}
q(t)&=&-\frac{\dot{H}}{H^2} -1\,, \nonumber\\
\,\nonumber\\
j(t)&=&\frac{\ddot{H}}{H^3}-3q-2\,, \\
\,\nonumber\\
s(t)&=&\frac{\dddot{H}}{H^4}+4j+3q(q+4)+6\,.\nonumber
\end{eqnarray}

\begin{figure}[h!]
\includegraphics*[scale=0.3]{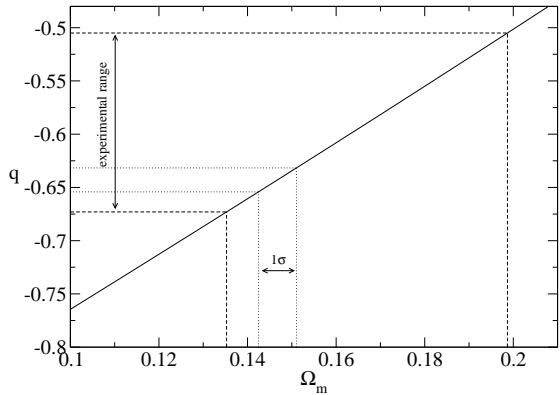}
\caption{Cosmological parameter $q$ as a function of $\Omega_m$ for
mod$_1$. The experimental interval of $q$ lies between the two
dashed lines, while the dotted lines delimit the $1\sigma$
confidence interval of $\Omega_m$ obtained from the analysis of the
expansion history.}\label{1q}
\end{figure}

\begin{figure}[h!]
\includegraphics*[scale=0.3]{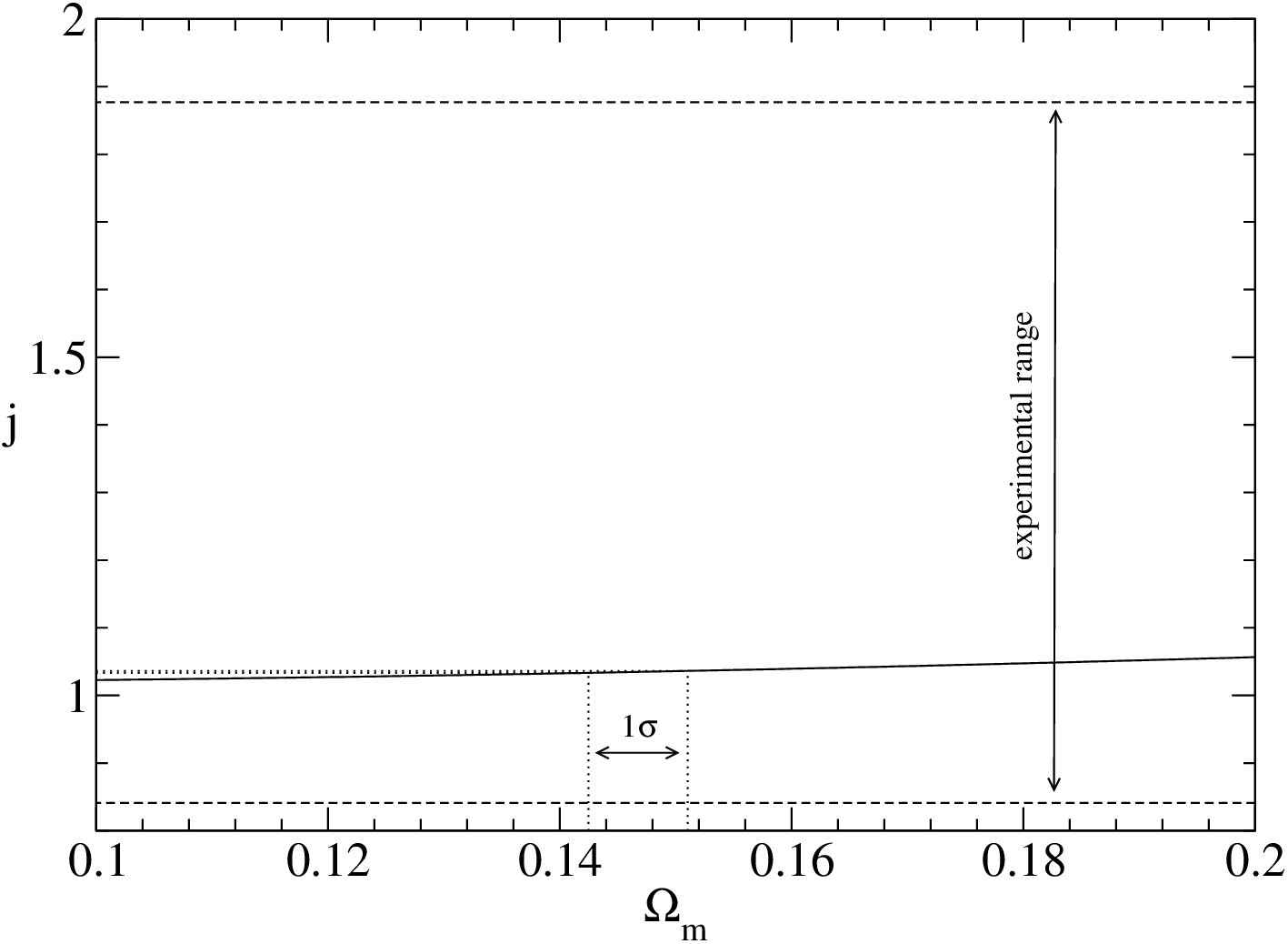}
\caption{Cosmological parameter $j$ as a function of $\Omega_m$ for
the model mod$_1$. The experimental interval of $q$ lies between the
two dashed lines, while the dotted lines delimit the $1\sigma$
confidence interval of $\Omega_m$ obtained from the analysis of the
expansion history.}\label{1j}
\end{figure}

\begin{figure}[h!]
\includegraphics*[scale=0.3]{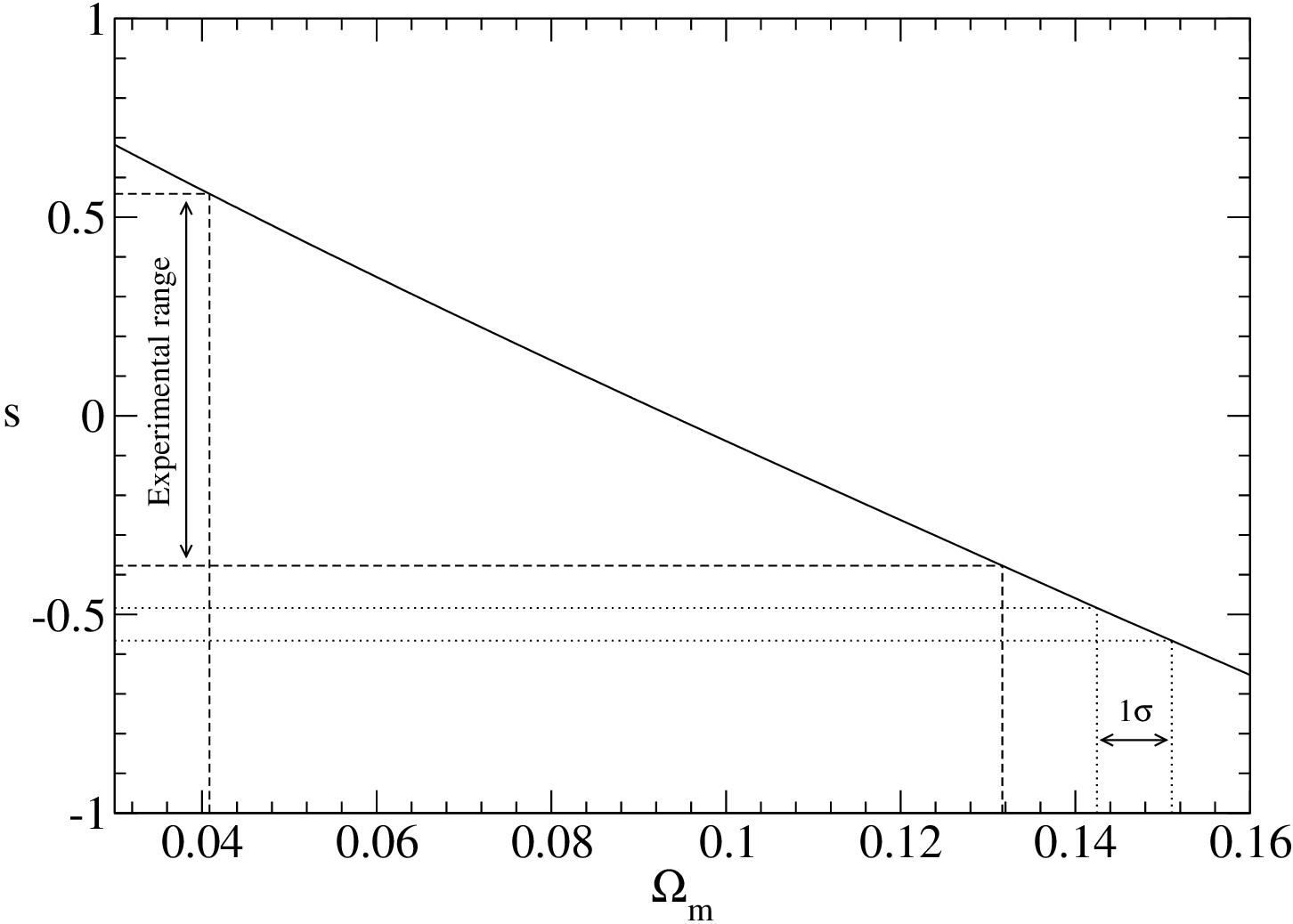}
\caption{Cosmological parameter $s$ as a function of $\Omega_m$ for
the model mod$_1$. The experimental interval of $q$ lies between the
two dashed lines, while the dotted lines delimit the $1\sigma$
confidence interval of $\Omega_m$ obtained from the analysis of the
expansion history.}\label{1s}
\end{figure}

For consistency, we fix, for both models, the values of $\alpha$ as
in the previous section, and we develop the results shown in Figs.
\ref{1q}, \ref{1j} and \ref{1s}, where we plot the values of $q$,
$j$ and $s$, respectively, at $z=0$ as functions of $\Omega_m$ for
$mod_{1;3}$. In each plot, the dashed lines delimit the experimental
range within the error bars and the corresponding interval of
$\Omega_m$, while the dotted lines delimit the $1\sigma$ interval of
confidence inferred from the expansion history analysis and the
corresponding range of values of the cosmological parameters.

Notice that the values of $q$ and $j$ obtained for $\Omega_m$ within
the $1\sigma$ confidence interval are in agreement with the
observations; only the parameter $s$ lies slightly outside the
experimental range. The experimental ranges of the cosmological
parameters and the values computed for $\Omega_m$ within the
$1\sigma$ confidence interval are written in tables \ref{tab1} and
\ref{tab2} for the $mod_1$ and $mod_3$, respectively. However, we
refer to the \emph{experimental ranges} as the numerical results
that we obtained by fitting directly Eq. ($\ref{zump}$) through the
use of Union 2, while the \emph{theoretical results} as the values
obtained by inverting the CS in terms of $\Omega_m$ using the values
found in Table I for each model;

\begin{table}[h!]
\begin{center}
\begin{tabular}{|c|c|c|c|}
\hline
Parameter  &  Experimental range & Theoretical range\\
\hline
q & -0.589$\pm$ 0.084 & $-0.654< q <-0.632$ \\
\hline
j & 1.359$\pm$ 0.518 & $1.033< j <1.036$ \\
\hline
s & 0.091$\pm$ 0.468 & $-0.566< s <-0.483$ \\
\hline
\end{tabular}
\caption{Experimental ranges of the cosmological parameters and
their values computed for $\Omega_m$ within the $1\sigma$ confidence
interval for $mod_1$ $0.143<\Omega_m <0.151$.} \label{tab1}
\end{center}
\end{table}

\begin{table}[h!]
\begin{center}
\begin{tabular}{|c|c|c|c|}
\hline
Parameter  &  Experimental range & Theoretical range\\
\hline
q & -0.589$\pm$ 0.084 & $-0.179< q <-0.160$ \\
\hline
j & 1.359$\pm$ 0.518 & $0.324< j <0.345$ \\
\hline
s & 0.091$\pm$ 0.468 & $-1.287< s <-1.212$ \\
\hline
\end{tabular}
\caption{Same as in table \ref{tab1}, but for $mod_3$ ($0.138<
\Omega_m <0.148$).} \label{tab2}
\end{center}
\end{table}

it follows from the two tables that both  models are compatible with
the experimental limits offered by Cosmography. Indeed, while
$mod_1$ excellently reproduces the experimental results, for $mod_3$
the values of the cosmological parameters $q$, $j$, $s$ at $z=0$,
obtained for $\Omega_m$ within the $1\sigma$ confidence interval,
are not at the same agreement level as $mod_1$. An accurate look at
the results shows that the goodness of $mod_3$ remains
disadvantaged, since it behaves worse than $mod_1$, which seems to
appear more predictive, but it should not be ruled out definitively
since the signs remain in the range of compatibility. Notice that
the goodness offered by the cosmographic test actually reflects the
intriguing physical aspect relying on the fact that in a low
redshift regime we should expect that $mod_1$ behaves like the
$\Lambda$CDM. In other words, we can expect that $mod_1$ formerly
reduces to a cosmological constant at small redshift more quickly
than $mod_3$.

\section{Inhomogeneities and Anisotropies of the CMB power spectrum}

In this section, we describe the imprint of anisotropies into the
CMB power spectrum, by adopting our models and we develop the growth
of inhomogeneities for the matter sector as well.

Since our models do not provide any analytical expressions for the
anisotropic equations, we are trying to map both $mod_1$ and $mod_3$
by finding a suitable approximation to the expansion history for
them. The basic idea is to map both the models, by using the results
obtained in Sec. II, in order to check if they work at higher
redshifts as well.

What we will immediately infer is that their behaviors suggest that
the EoS can be approximated by three separated pieces, a dark matter
fluid, a relativistic fluid and a cosmological constant term.

It turns out to be more accurate to perform this approximation,
giving rise to an evolving EoS, precisely we follow the standard
Chevallier-Polarski-Linder (CPL) parametrization \cite{CPLCP,CPLL},
which invokes as a barotropic factor the known expression $w=w_0 +
w_a\frac{z}{1+z}$.

Thus, we approximate the energy density of the holographic fluids
($\rho_X$) as

\begin{eqnarray}
\frac{8 \pi G}{3 H^2_0} \rho_X &\approx& \Omega_{Xdm}(1+z)^3 +
\Omega_{Xr}(1+z)^4  \label{approx} \\ \nonumber & + & \Omega_{CPL}
(1+z)^{3(1+w_0 + w_a)} \exp\left(-3w_a \frac{z}{1+z} \right).
\end{eqnarray}

Here, the subscript $Xdm$ stands for the \emph{holographic DM},
mimicking the piece behaving as matter in the holographic EoS, while
$Xr$ analogously represents the relativistic part, which mimics the
relativistic term as well as the matter.

For purposes of CMB analysis we must demand a good approximation at
early times, before recombination. A failure in the approximation at
late times will be reflected in the large scale multipole moments
($low-l$) which are poorly constrained due to the cosmic variance.

For both models we adopt $\Omega_r h^2 =2.469 \times 10^{-5} $ and
$h=0.704$ \cite{WMAP7y} and regarding $\Omega_m$, we assume the
validity of the best fits given by SNeIa and CMB shift, found in
Sec. III for each model.

\begin{figure}
\centering
\includegraphics[width=3.4in]{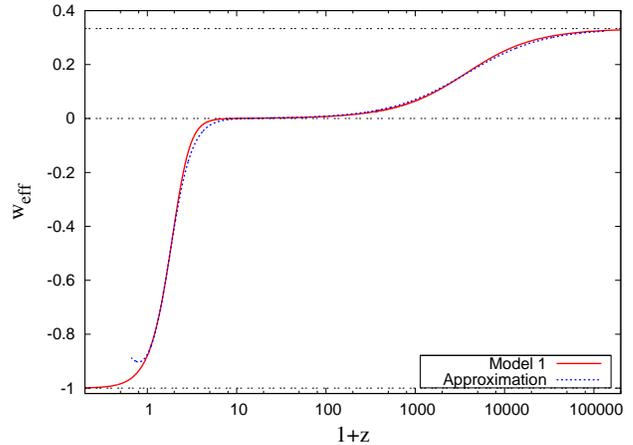}
\caption{(Color online). Effective EoS parameter of the holographic
fluid for model 1 and for the approximation we have made from it
using Eq. \ref{approx}.} \label{fig5}
\end{figure}

\begin{figure}
\centering
\includegraphics[width=3.4in]{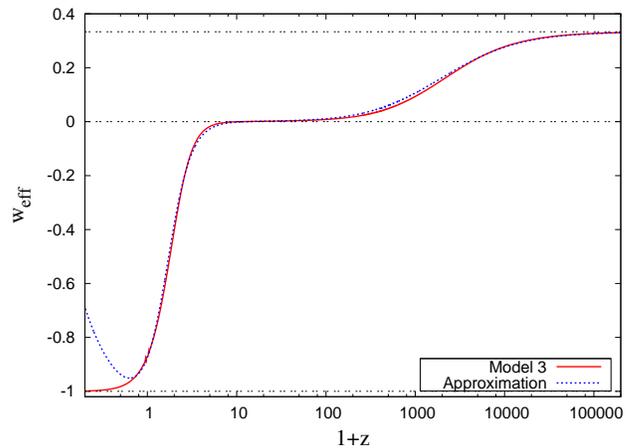}
\caption{(Color online). The same as Fig. \ref{fig5}, but for model
2.} \label{fig6}
\end{figure}

The numerical analysis shows that the compatible results are given
by having for mod$_1$: $w_0=-1.04$, $w_a=-0.2$,
$\Omega_{Xdm}=0.139$, $\Omega_{Xr} = 0.72\, \Omega_{r} $, and
$\Omega_{CPL}$ fixed by the flat condition, e.g. $\Omega_{Xdm} +
\Omega_{Xr}+\Omega_{CPL}+\Omega_{m}+\Omega_{r}=1$. In Fig.
\ref{fig5} we plot the effective EoS parameter of mod$_1$ under our
approximation. It is well emphasized that the differences are
extremely small, giving rise to good results of the first model at
higher redshift. In particular, they have been estimated to be less
than $2\%$, before recombination ($z \sim 1100$).

For mod$_3$ analogously we have found $w_0=-1.05$, $w_a=-0.09,$
$\Omega_{Xdm}=0.135$ and $\Omega_{Xr} = 0.7\, \Omega_{r} $.
$\Omega_{CPL}$ is again fixed by the requirement of spatially flat
geometry like the above case. Figure 9 shows the effective EoS
(weff) and the approximation based on the CPL parametrization. In
this case, the differences are also small, being less than $3\%$
before recombination.

It appears intriguing to note that these approximations are not
valid for future redshift, i.e. $z<0$; they cannot be extrapolated
to the future.

Figures. \ref{fig5} and \ref{fig3} show the EoS parameters of
mod$_{1;3}$, respectively.

Moreover, for both models the relativistic term behaves in the form
$\Omega_{Xr} \sim 0.7 \, \Omega_{r}$, while for nonrelativistic
matter we found $\Omega_{Xdm} \sim 0.95 \,\Omega_m$. This
definitively shows that the holographic models enhance the
gravitational effects due to nonrelativistic fluids more than that
of relativistic ones; we are not surprised by this feature, since,
though in different ways, we have already anticipated it in Sec. III

\begin{figure}
\centering
\includegraphics[width=3.4in]{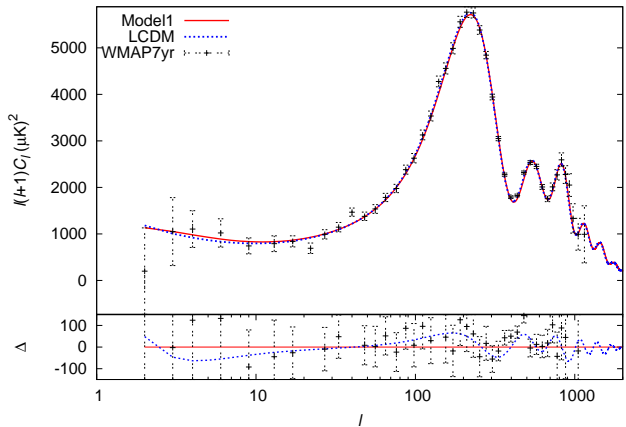}
\caption{(Color online).CMB TT power spectrum for model 1 (thick
line) and the LCDM model (dotted line). The error bars refer to the
         binned results of the WMAP 7 \cite{WMAP7y}. The difference between model 1 and the LCDM model ($\Delta$) is also shown.}
\label{fig7}
\end{figure}

\begin{figure}
\centering
\includegraphics[width=3.4in]{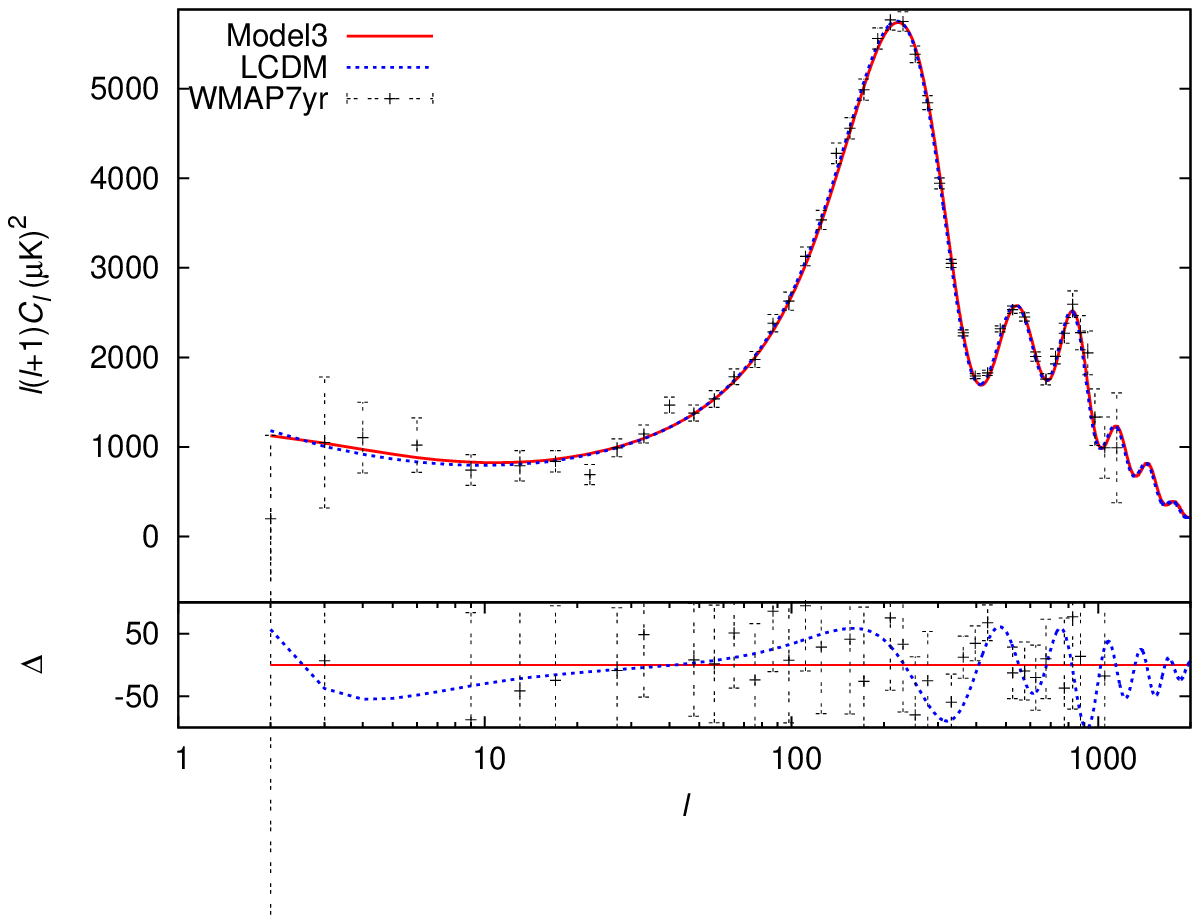}
\caption{(Color online). The same as Fig. \ref{fig7} but for model
2.} \label{fig8}
\end{figure}

\begin{figure}
\centering
\includegraphics[width=3.4in]{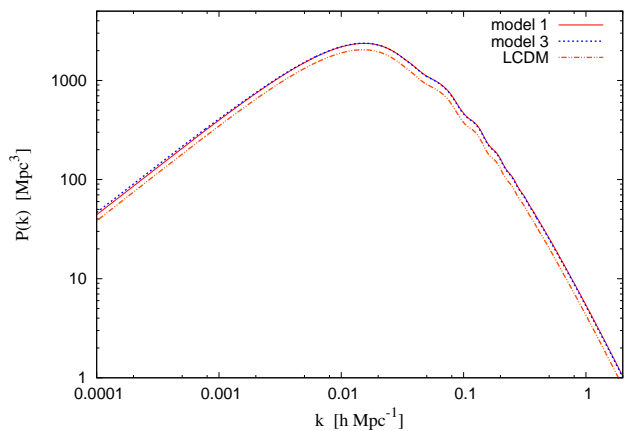}
\caption{(Color online).Matter power spectrum for model 1, model 2
and the LCDM model.} \label{fig9}
\end{figure}

In order to perform the power spectrum analysis, we proposed the use
of the publicly available code CAMB \cite{Lewis00}; in particular,
we compare the holographic models with respect to the observational
data and the $\Lambda$CDM model as well.

Figures \ref{fig7} and Fig. \ref{fig8} give the experimental
results. In those figures, we plot the CMB TT angular power spectrum
and the residual,
\begin{equation}\label{resu}
\Delta \equiv l(l+1) (C_l^{\Lambda\text{CDM}}-C_l^{model})\,.
\end{equation}
We elucidate great differences concentrated at large scales, which
can be justified since the systematic errors due to the cosmic
variance are strongly dominant. In addition, there are large
discrepancies in the first three peaks, but these have been inferred
to be well inside the error bars, different from the rest. 
At low scales, those differences quickly decay; this is as expected,
because in the Silk damped tail region the anisotropies are mainly
due to the microphysics driven by the photon-baryon plasma which the
holographic fluid is not able to modify; indeed, the holographic
fluids interact only gravitationally.

In order to obtain the plots of Figs. \ref{fig7} and \ref{fig8}, we
have also fixed the amplitude of the primordial scalar perturbation
$A_s$, defined as the proportionally constant in the equation
\begin{equation}\label{qjanz}
k^3 P_{\Phi}(k) \propto (k/k_0)^{n_s -1}\,,
\end{equation}
where $P_{\Phi}$ is the primordial power spectrum of the
gravitational potential $\Phi$, while $n_s$, defined as the spectral
index, is assumed to be $n_s\simeq 0.967$ and $k_0 = 2\times10^{-3}
\text{Mpc}^{-1}$, where we moreover put an arbitrary pivotal scale
$k$ in the above expression. For both holographic models and for
$\Lambda$CDM, which is used for calibrating our tests, we adopted
respectively $A_{s\,mod1}=2.48 \times 10^{-9}, A_{s\,mod3} = 2.42 \times
10^{-9}$.

On the other hand, we emphasize that Fig. \ref{fig9} shows the
matter power spectrum of mod$_{1;3}$; we found extremely small
discrepancies between the three curves, where the third one is
represented by $\Lambda$CDM. The larger differences are located
about $k\sim 0.1\,h \text{Mpc}^{-1}$, leaving probable imprints in
the baryon acoustic oscillations that could be potentially detected
in near future experiments. Nonetheless, currently these differences
are not enough to discriminate between the $\Lambda$CDM model and
one of our models. This confirms that there is no reason not to use
against the use of curvature invariants as a tool for explaining
both DM and DE effects.

\section{conclusion}

In this work, we proposed as IR cutoff scale for the size $L$, in
the context of the cosmological HP, a second order geometrical
approach, dealing with the use of independent invariants embedded in
a FRW background, as a source for both DM and DE.

In particular, the DE density is assumed to be proportional to these
invariants, which reduce from 14 to 3 in the FRW cosmology; thus, we
use only GR and the HP to construct our models. Moreover, we
extended the work of \cite{mio} and we safely overcame the problem
of causality, portrayed in \cite{cai}, which represents one of the
most serious shortcomings of the choice of the IR cutoff.

One of the main benefit of our approaches is basically due to the
advantage of characterizing the acceleration of the Universe, by
geometrical considerations only, by pointing out that the geometry
is capable to describe the positive acceleration.

Particularly, we found two viable models (the third one is trivial,
corresponding to the dustlike case only), and we developed a series
of cosmological tests, able to explain their robustness. Therefore,
the expansion history definitively fixed the values of $\Omega_m$
and $\alpha$, the free parameters of our approaches, which have been
described through the use of SNeIa and CMB tests.

Moreover, as a powerful analysis, we investigated cosmography in the
framework of our models, since it has been testified an amusing
interest on the Universe's kinematics; the predictions of
cosmography certified that $mod_1$ is favored if compared to
$mod_3$, as it has been pointed in Sec. III. We also arrived  at
analogous results in the last sections, with higher redshift tests.
In fact, additional confirmation came from the direct study of the
anisotropy of the power spectrum and then, after all, we conclude
that our models can be pondered as possible sources of both DE and
DM, becoming a viable candidate to explain unified schemes for the
Universe's dynamics; in particular, such a picture seems to be able
to naturally reduce the problem of the DM presence in the Universe,
in the framework of GR, being able to explain, at the same time, the
cosmic speed up.

More considerations can be carried forward this idea, in order to
check if, trough higher order invariants, it would be possible to
definitively overcome the problem of DM and DE. This concept will be
the object of future developments.

\section*{Acknowledgements}

It is a pleasure to thank Dr. Andrea Geralico and Professor
Salvatore Capozziello for very fruitful discussions.


\begin{thebibliography}{99}

\bibitem{cliff}
M. W. Clifford, Physics, 4, 43, (2011).

\bibitem{SNeIa1}
A. G. Riess et al., AJ, 116, 1009, (1998)

\bibitem{SNeIa2}
S. Perlmutter et al., ApJ, 517, 565, (1999).

\bibitem{sn3}
R. Rebolo et al., MNRAS, 353, 747, (2004); A. C. Pope et al., ApJ,
607, 655, (2004); P. McDonald et al., astro-ph/0405013, (2004); M.
Tegmark et al. (SDSS), Phys. Rev. D74, 123507 (2006); W. J. Percival
et al., Mon. Not. Roy. Astron. Soc. 381,1053, (2007).

\bibitem{coppa}
E. J. Copeland, M. Sami, S. Tsujikawa, Int. J. Mod. Phys. D, 15,
1753-1936, (2006).

\bibitem{lam}
S. Weinberg, Cosmology, Oxford Univ. Press, Oxford, (2008); S.
Weinberg, Rev. Mod. Phys. 61, 1, (1989).

\bibitem{tsu}
S. Tsujikawa, ArXiv: 1004.1493.

\bibitem{nu}
M. Li, X. D. Li, S. Wang, Y. Wang, ArXiv: 1103.5870.

\bibitem{hol}
P. McFadden, K. Skenderis, Phys. Rev. D 81, 021301(R), (2010); R.
Bousso, Rev. Mod. Phys.74:825-874, (2002).

\bibitem{bingo}
Z. Chang, F. Q. Wu and X. Zhang, Phys.\ Lett.\  B {\bf 633}, 14,
(2006); B. Wang, C.Y. Lin and E. Abdalla, Phys. Lett. B {\bf 637},
357, (2006).

\bibitem{uhmamma2}
X. Zhang, Phys.Lett.,  B {\bf 648}, 1, (2007); M. R. Setare, J.
Zhang and X. Zhang, JCAP, {\bf 0703}, 007, (2007); J. Zhang, X.
Zhang and H. Liu, Phys.\ Lett.\  B {\bf 659}, 26, (2008).

\bibitem{uhmamma3}
C.J. Feng, Phys. Lett. B, {\bf 633}, 367, (2008);M. Li, X.D. Li, C.
Lin and Y. Wang, Commun. Theor. Phys., {\bf 51}, 181, (2009); M.
Jamil, M. U. Farooq and M. A. Rashid, Eur. Phys. J. C, {\bf 61},
471, (2009).

\bibitem{mio}
L. Bonanno, G. Iannone, O. Luongo, ArXiv: 1101.5798.

\bibitem{cai}
R. G. Cai, Phys. Lett. B {\bf 657}, 228, (2007).



\bibitem{capozzox}
C. Cherubini, D. Bini, S. Capozziello, R. Ruffini, Int. J. Mod.
Phys. D,11, 827-841, (2002).

\bibitem{roberts}
M. D. Roberts, Int. J. Mod. Phys., 9, 167, (1994).

\bibitem{gendeb}
J. G\'eh\'eniau and R. Debever, Bull. Cl. Sci. Acad. R. Belg. XLII,
114, (1956).

\bibitem{Witten}
L. Witten, Phys. Rev., 113, 357, (1959).

\bibitem{Petrov}
A. Z. Petrov, Eistein Spaces, Pergamon, Oxford, (1969).

\bibitem{rinpen}
R. Penrose and W. Rindler, Spinors and Spacetime, Cambridge Univ.
Press, (1986).

\bibitem{carmi}
J. Carminati and R. G. McLenaghan, J. Math. Phys., 32, 3134, (1991).

\bibitem{WMAP7y} Komatsu {\it et al.}, Astrophys. J. Suppl. Ser. {\bf 192}, 18 (2011).
                 Larson {\it et al.}, Astrophys. J. Suppl. Ser. {\bf 192}, 16 (2011).

\bibitem{Union2}  R. Amanullah {\it et al.}, Astrophys. J.  {\bf 716}, 712 (2010).

\bibitem{EstaBond} J.R. Bond, G. Efstathiou and M. Tegmark, MNRAS {\bf 291} L33 (1997)

\bibitem{Melch} A. Melchiorri and  L. M. Griffiths,  New Astron.Rev. {\bf 45}  321 (2001)

\bibitem{CPLCP} M. Chevalier and D. Polarski, Int. J. Mod. Phys. D., 10, 213,(2001).

\bibitem{CPLL} E. Linder, Phys. Rev. Lett., 90, 091301, (2003).

\bibitem{Lewis00} A. Lewis, A. Challinor, and A. Lasenby, Astrophys. J {\it 538}, 473 (2000).




\end{thebibliography}
\end{document}